\documentclass[letterpaper,12pt]{article}   
\usepackage{osajnl2} 

\usepackage{latexsym}
\usepackage[T1]{fontenc}
\usepackage[latin9]{inputenc}
\usepackage{float}
\usepackage{array}
\usepackage{amsmath}
\usepackage{amssymb}
\usepackage{graphicx}
\usepackage{esint}
\begin{document}

\title{Compensation of the laser parameters fluctuations in large ring laser gyros: a Kalman filter approach}

\textit{Applied Optics}


\author{Alessandro Beghi$^{1}$, Jacopo~Belfi$^{2*}$, Nicol\`o~Beverini$^{2}$, B. Bouhadef$^{2}$, D. Cuccato$^{1,3}$, Angela Di Virgilio$^{4}$, 
 and  Antonello Ortolan$^{3}$}
\address{$^1$Department of Information Engineering, University of Padova
Via Gradenigo 6/B, Padova, Italy}
\address{$^2$Department of Physics ``Enrico Fermi,'' Universit\`a di Pisa,
and\\ CNISM unit\`a di Pisa, Italy}
\address{$^3$INFN National Laboratories of Legnaro, Viale dell'Universit\`a
2, Legnaro, Padova, Italy}
\address{$^4$INFN Sez. di Pisa, Pisa, Italy}
\address{$^*$Corresponding author: belfi@df.unipi.it}

\begin{abstract}
He-Ne ring laser gyroscopes are, at present, the most precise devices for absolute angular velocity measurements. Limitations to their performance 
come from the non--linear dynamics of the laser. Following the Lamb semi-classical theory, we find a set of critical parameters affecting the time 
stability of the system. We propose a method for estimating the long term drift of the laser parameters and for filtering out the laser dynamics 
effects from the rotation measurement. The parameter estimation procedure, based on the perturbative solutions of the laser dynamics, 
allow us to apply Kalman Filter theory for the estimation of the angular velocity. Results of a 
comprehensive Monte Carlo simulation and results of a preliminary analysis on experimental data from the ring laser 
prototype G-Pisa are shown and discussed. 
\end{abstract}

\section{Introduction}

Ring laser ``gyros'' are standard sensors in estimating rotation rates relative to the inertial frame, 
with many applications ranging from inertial guidance \cite{inertialguidance}, to angle metrology \cite{anglemetrology}, 
geodesy \cite{geodesy, chandler}, geophysics \cite{geophysics}, as well to fundamental physics \cite{geophysics,G-PISA}. 
In the near future their application is foreseen  to improve the performances of advanced gravitational waves detectors 
\cite{gravwavedetector} and  also to provide ground based tests of General Relativity \cite{ginger}.

In a ring laser two oppositely traveling optical waves resonate inside a polygonal closed path. 
In an inertial frame, each beam follows a path of the same length. When the cavity rotates, a round--trip 
time difference between the co-rotating and counter-rotating beams occurs, as they experience a longer and a shorter path, 
respectively. This translates into a frequency difference of the two beams. Such frequency difference contains the information 
about the rotation rate of the reference frame. 

This physical phenomenon is known as Sagnac effect and the Sagnac frequency $\nu_{s}$ (i.e. the frequency of the beat signal between the two beams), reads 
\begin{equation}
\nu_{s}=\frac{4A}{\lambda L}\,\mathbf{n}\cdot\mathbf{\Omega},\label{eq:stmodel1}
\end{equation}
where $A$ and $L$ are the area and the perimeter of the cavity respectively, $\lambda$ is the wavelength of the laser beam, 
$\mathbf{n}$ is the normal vector of the plane of the ring cavity, $\mathbf{\Omega}$ is the rotation rate vector. 
In a real ring laser the laser dynamics are source of systematic errors in the estimate of $\nu_s$. 
The laser dynamics are determined by a set of non linear equations that depend on parameters that are slowly varying, following random changes of the environmental conditions (mainly temperature and atmospheric pressure). In addition, the control of some parameters (e.g. laser emission frequency or light intensity) in a closed loop way may enhance the drifts of the others. 
Thus the simple Eq.(\ref{eq:stmodel1})  must be modified to take into account such complex behaviors (see for example \cite{geophysics}) which 
result in  \emph{i}) corrections to scale factor due to fluctuations of  the laser gain, cavity losses and frequency detuning; \emph{ii}) null shift due to any cavity non-reciprocity; and 
\emph{iii}) non linear coupling between the two laser beams due to backscattering (at present, the most 
important instability source). Therefore, the beating signal can be represented as the sum of $\nu_s$ , 
white noise and terms arising from the non linear dynamics. Usually, the last ones are removed by long term correlations with a set of auxiliary sensors information, and by modeling the ring laser behavior 
with respect to the ideal case \cite{Alex}. However, the non linearity of the dynamics limits the effectiveness of this approach. In this paper we propose to increase the ultimate resolution and the time stability of ring lasers by the estimation of the parameters of their system, and the subsequent application of an Extended Kalman filter (EKF) \cite{EKF}. 
 
The proposed method has a very wide range of application. In fact, the parameter estimation and the dynamical filtering could improve the response 
of large size ring lasers, which have already reached accuracy of $\sim 1$  part in $10^{9}$, and so providing unique informations for geophysics and geodesy. 
Middle size rings, with sides of $\sim 1$ m, which are more affected by backscattering, will be improved as well. Such instruments are more suitable for 
geophysics applications (i.e. rotational seismology), and for application to gravitational waves interferometers (e.g. local tilts measurements).

The paper is organized as follows. In Section \ref{sect2} we discuss the non--linear dynamics of a 
He-Ne inhomogenously broadened ring laser. Here the critical parameters affecting ring behavior are presented. 
Section \ref{sect3} describes the implementation of the parameter estimation procedure and the Extended Kalman Filter (EKF) algorithm \cite{EKF}. 
In Section \ref{sect4}  we presents the results of the parameter estimation procedure, and the performance of EKF for rotation estimation. 
The procedure has been tested on the experimental data of G-Pisa, and results are reported in Section \ref{sect5}. 
Finally, our conclusions are drawn in Section \ref{sect6}. 

\section{ \label{sect2} Dynamics of a Ring Laser} 
The differential equations of the ring laser dynamics were first derived long ago, using a low 
gain self consistent model by E. Lamb \cite{Lamb} for linear laser, and then extended to rings by F. Aronowitz \cite{Aronowitz1}
\begin{align}
{\displaystyle \dot{I}_{1}\,=} & \,\frac{c}{L}\left[\alpha_{1}I_{1}-\beta_{1}I_{1}^{2}-\theta_{12}I_{1}I_{2}+2r_{2}\sqrt{I_{1}I_{2}}\cos(\psi+\varepsilon)\right]\nonumber \\
\dot{I}_{2}\,= & \,\frac{c}{L}\left[\alpha_{2}I_{2}-\beta_{2}I_{2}^{2}-\theta_{21}I_{1}I_{2}+2r_{1}\sqrt{I_{1}I_{2}}\cos(\psi-\varepsilon)\right]
\label{eq:fond1}\\
\dot{\psi}\,= & \,\omega_{s}+\sigma_{2}-\sigma_{1}+\tau_{21}I_{1}-\tau_{12}I_{2}-\nonumber \\
 & -\frac{c}{L}\left[r_{1}\sqrt{\frac{I_{1}}{I_{2}}}\sin(\psi-\varepsilon)+r_{2}\sqrt{\frac{I_{2}}{I_{1}}}\sin(\psi+\varepsilon)\right]\ ,\nonumber 
\end{align}
where $I_{1,2},$ $\psi$ and $\dot{\psi}$ are the dimensionless light intensities, the instantaneous phase difference, 
and the instantaneous circular beat frequency of the counter-propagating waves, respectively. 
Here $\omega_{s} = 2 \pi \nu_s $ is the rotation rate in Eq.(\ref{eq:stmodel1}), and $\alpha_{1,2}, \sigma_{1,2}, \beta_{1,2}, \theta_{12,21},  \tau_{12,21}, r_{1,2}, \varepsilon$ are the Lamb parameters. It is worth mentioning that  $\alpha_{1,2}, \sigma_{1,2}$  are the amplification minus losses, $  \beta_{1,2} $ is the self saturation, $  \theta_{12,21},  \tau_{12,21} $  describe cross-(mutual-)saturation, $  r_{1,2}, \varepsilon$  are the amplitude and the relative phase of the backscattered waves, respectively. A more detailed explanation of the physical meaning of these parameters can be found in Appendix \ref{sec:Lamb-Coefficient-Calculation}.
\subsection{Lamb parameters effects on the gyroscope performances} 
The study of Eqs.(\ref{eq:fond1}) has been conducted by several authors in the past \cite{Aronowitz2,Stedman,Mandel,PBR,Mandel2}, 
  both with numerical and analytical approaches. The analytical solution cannot be found in the most general case, but only under certain approximations 
about the reciprocity of the system. Approximated analytical expressions for the time evolution of the Sagnac phase provide, nevertheless, an useful
reference to better understand the role of the Lamb parameters noise on the estimation of the angular velocity $\omega_s$. 

We present in the following the periodic solution of Eqs.(\ref{eq:fond1}) in the case where: $I_{1}/I_{2}=k$ and $\tau_{1 2}=\tau_{2 1}=0$. 
In this case the phase equation takes the form:  
\begin{equation}
\dot{\psi}\,=\omega_{s}-\frac{c}{L}\left[r_{1}k\sin(\psi-\varepsilon)+\frac{r_{2}}{k}\sin(\psi+\varepsilon)\right]\quad,
\label{eq:Adler}
\end{equation}
and admits the following solution:
\begin{equation}\label{eq:SolAdler}
\psi(t)={\displaystyle 2\arctan\left[\frac{\Omega_{L1}+\Omega_{p}\tan\left(\frac{1}{2}\Omega_{p}t\right)}{\omega_{s}+\Omega_{L2}}\right]\quad,}
\end{equation}

where $\Omega_{L1}=c/L(kr_{1}+r_{2}/k)\cos(\varepsilon)$, $\Omega_{L2}=c/L(r_{2}/k- kr_{1}) \sin(\varepsilon)$ 
and $\Omega_{p}=\sqrt{\omega_{s}^{2}-(\Omega_{L1}^{2}+\Omega_{L2}^{2})}$.

From Eq. (\ref{eq:SolAdler}), for $\Omega_{L1,L2}\ll\omega_{s}$, we get
\begin{equation}
\omega(t)\,\simeq\,\omega_{s}-\Omega_{L2}\cos\omega_{s}t-\Omega_{L1}\sin\omega_{s}t\quad,\label{eq:ErrorModel}
\end{equation}

where $\omega$ denotes the detected Sagnac frequency and 
$ \omega_{BS} \equiv - \Omega_{L2} \cos\omega_{s}t - \Omega_{L1} \sin\omega_{s}t$ 
represents the frequency modulation of the Sagnac signal. 

\begin{figure}
\centering
\includegraphics[width=14cm]{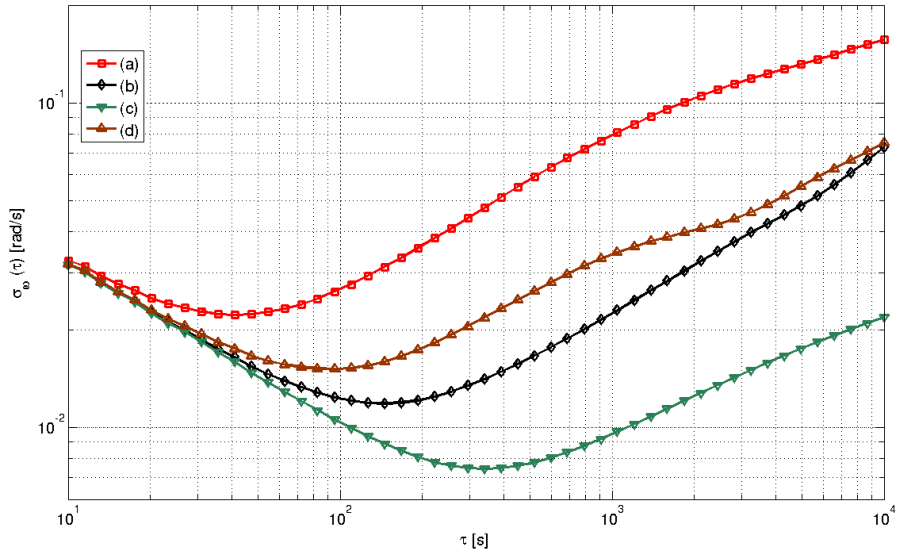}
\caption{Simulated Allan deviations of the estimated rotation rate. 
See the text for details.}
\label{fig:Montecarlo_simulation_AVAR}
\end{figure}

In Figure \ref{fig:Montecarlo_simulation_AVAR} we report the results of a Monte Carlo simulation of $10^{6}\rm{s}$ of ring laser dynamics 
evolution with $I_{1}/I_{2}=k$. 
We considered the following noise sources in the system: a white frequency noise with standard deviation of $10^{-1}$ on $\omega(t)$
 (mimic of the output of AR(2) frequency detection algorithm \cite{AR2}) and a random-walk noise
on the parameters $r_{1,2},\, k,$ and $\varepsilon$. 
The Allan variance of $\omega(t)$ has been calculated for four different cases, denoted with (a), (b), (c) and (d).

In case (a) $(r_{1,2}, k,\varepsilon)$ are independent and vary in random walk with a step size of $(2\cdot 10^{-9}, 10^{-3}, 10^{-2})$ respectively. 
In case (b) the only varying parameter is $k$, with a step size of $10^{-3}$. 
In case (c) all parameters vary as in (a), but the processes $r_{1}$ and $r_{2}$ have been correlated  with a correlation 
coefficient of $0.9$ while $\varepsilon$ varies around the nominal value of $0\, \rm{rad}$ with a random walk 
step size of $10^{-4}\, \rm{rad}$. 
In case (d) all parameters vary as in (b), but the process $\varepsilon$ 
varies around the nominal value of $\pi/2\, \rm{rad}$ with a random walk step size of $10^{-4}\, \rm{rad}$.  

It can be easily observed that the noise contribution coming from the parameters fluctuation is transferred to the 
noise of the measured Sagnac frequency exhibiting the same random walk plus white noise pattern. 
The relative noise on the laser parameters is converted into frequency noise by the factor $c/L$  
meaning that the larger is the cavity perimeter, the larger is the rejection of the laser parameters noise. 
In addition, it is worth noticing that the backscattering phase $\varepsilon$ plays a crucial role in 
transfering the fluctuations of $r_1,r_2$ on $\omega$. It determines a strong reduction of the output 
noise for values close to $\varepsilon=\pi/2$ (trace (d)). In this regime, also known as 'conservative 
coupling regime' \cite{Stedman}, the backscattered photons interact destructively and their influence on the 
nonlinear interaction between the two intracavity beams and the active medium is minimized.

In the next sections  we will show that a statistical filtering procedure is able to identify the Lamb parameters, and  to remove their slow drifts by using extended Kalman filtering. Thus the maximum resolution 
(i.e. the minimum value of  $\sigma^2_\omega(\tau)$) and the time stability 
(i.e. the  value of $\tau$ where the minimum is attained) of a ring laser can be significantly improved. 
In fact,  the  instantaneous Sagnac frequency will depend in general on the full system 
state $\left[I_{1}(t),\ I_{2}(t),\ \psi(t)\right]$, and so dynamic Kalman filtering can be more effective 
in estimating  $\omega_s$ than other approaches that rely on $\psi(t)$ only (e. g. the standard $AR(2)$ method for frequency estimation \cite{AR2}).  

\section{Dynamics of G-Pisa}
\begin{figure} 
\centering
\includegraphics[width=14cm]{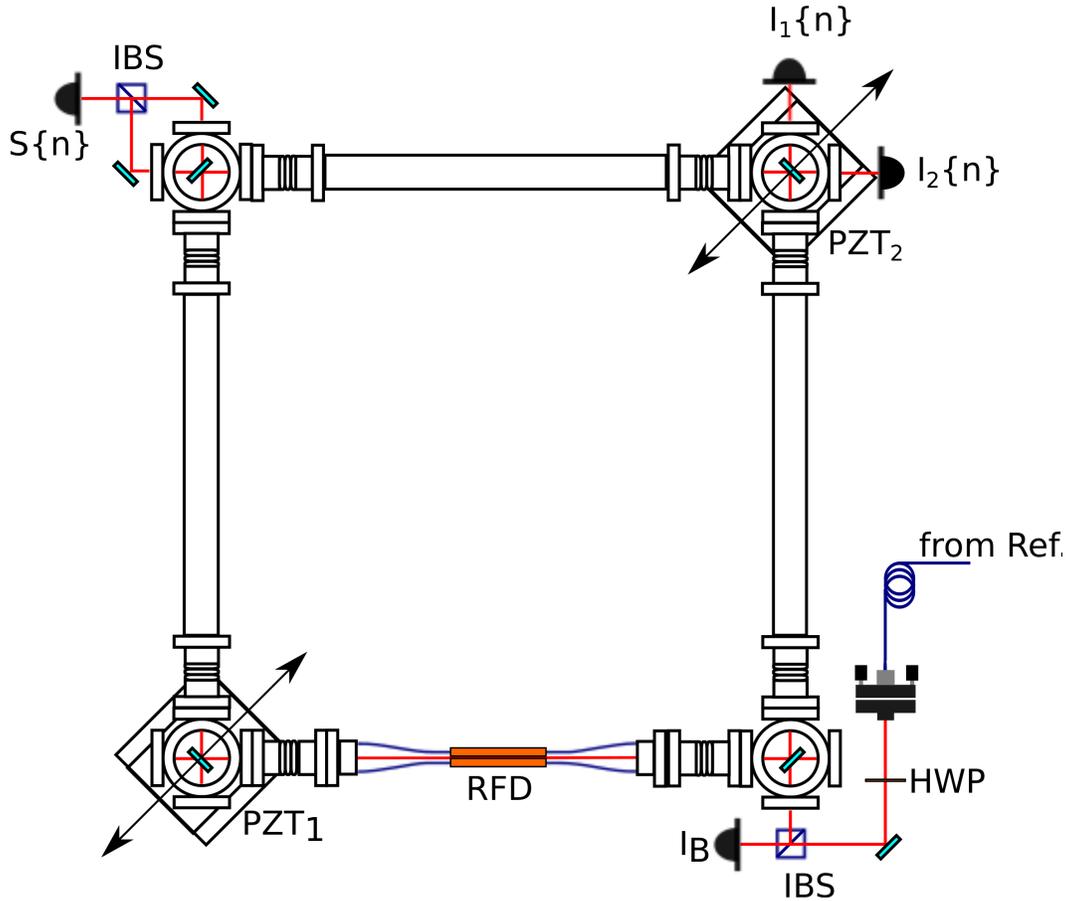}
\caption{G-Pisa experimental setup. The cavity vacuum chamber is entirely filled with a mixture of He-Ne and does not contain any intra-cavity element 
except for the four mirrors.  $\rm{S_\{n\}}$: Sagnac interference signal; $\rm{I_{1}\{n\}}$: counter-clockwise single beam intensity; 
$\rm{I_{2}\{n\}}$: clockwise single beam intensity; $\rm{I_B}$: optical beat intensity; RFD: radio frequency discharge; 
IBS: intensity beam splitter; HWP: half wave-plate; PZT: piezoelectric transducer.}
\label{setup}
\end{figure}

G-Pisa is a prototype middle size He-Ne ring-laser. 
The main characteristics of its optical cavity are reported in table \ref{tab_optic} while its experimental setup is sketched in figure \ref{setup}.
\begin{table}[h]
\centering
\begin{tabular}{|l|r|}
\hline
\multicolumn{2}{|c|}{\textbf{G-Pisa}}\\
\hline
\multicolumn{2}{|c|}{\it{Geometry}}\\
\hline
Cavity & square\\
\hline
Side length & 1.35 m\\
\hline
Latitude & 43$^\circ$ 40' 35.86"N\\
\hline
\multicolumn{2}{|c|}{\it{Cavity mirrors}}\\
\hline
Radius of curvature & 4 m\\
\hline
Total losses & 3.7 ppm\\
\hline
Transmission & 0.25 ppm\\
\hline
Scatter+absorption & 3.5 ppm\\
\hline
\multicolumn{2}{|c|}{\it{Optical properties}}\\
\hline
Wavelength & 632.8 nm\\
\hline
Output power & 1.6 nW (single mode)\\
\hline
Spatial mode & $\rm{TEM}_{00}$\\
\hline
Beam waist (s,h) & (1.97 mm, 2.43 mm)\\
\hline
\end{tabular}
\caption[Tab.1]{Main nominal characteristics of the ``G-Pisa" apparatus. Mirrors characteristics are the nominal ones (manufacturer information) and refer 
 to the single mirror for an incidence angle of 45$°$, and s-polarized light at 632.8 nm. Beam waists are reported as 4 times the $\rm{1/e^{2}}$ beam intensity 
radius.}
\label{tab_optic}
\end{table}

The laser operation is controlled by two feedback loop systems,  dedicated to the active stabilization of the optical frequency  
and the optical power \cite{G-PISA}. The first loop keeps constant the frequency difference between the ring laser 
clockwise beam and a reference laser. It acts on two piezoelectric transducers ($\rm{PZT_{1,2}}$) moving two 
opposite cavity mirrors along the cavity diagonal. The second loop regulates the RF discharge power in order to keep constant the power of 
the clockwise ring laser output.

G-Pisa dynamics can be derived from Eqs.(\ref{eq:fond1}) taking into account the spectroscopic properties of its active medium and the 
constraints imposed by the stabilization loops. The use of a special gas mixture, containing $\ ^{20}Ne$ and $\ ^{22}Ne$ isotopes at $50$:$50$ ratio,  
avoids mode competition effects, while the perimeter control, by keeping constant the laser optical frequency avoids mode-jumps. In the Appendix  \ref{sec:Lamb-Coefficient-Calculation} we show that  $\beta_{1}=\beta_{2}=\beta,\ \sigma_{1}=\sigma_{2},\ \tau_{12}=\tau_{21}=0$, and $\theta_{12}=\theta_{21}=0$ hold for the closed loop operation of G-Pisa. Thus the equations of the dynamics reduced to:

\begin{align}
{\displaystyle \dot{I}_{1}\,=} & \,\frac{c}{L}\left[\alpha_{1}I_{1}-\beta I_{1}^{2}+2r_{2}\sqrt{I_{1}I_{2}}\cos(\psi+\varepsilon)\right]\nonumber \\
\dot{I}_{2}\,= & \,\frac{c}{L}\left[\alpha_{2}I_{2}-\beta I_{2}^{2}+2r_{1}\sqrt{I_{1}I_{2}}\cos(\psi-\varepsilon)\right]\label{eq:GP}\\
\dot{\psi}\,= & \,\omega_{s} -\frac{c}{L}\left[r_{1}\sqrt{\frac{I_{1}}{I_{2}}}\sin(\psi-\varepsilon)+r_{2}\sqrt{\frac{I_{2}}{I_{1}}}\sin(\psi+\varepsilon)\right]\ .\nonumber 
\end{align}

In Table \ref{RefPar} we report the typical Lamb parameters of G-Pisa, with a typical round-trip gain of $G=3 \times 10^{-5}$. It is worth noticing that, as the perimeter is kept constant by moving the ring mirrors, the backscattering phase angle $\varepsilon$ is not fixed, but it ranges from $0$ (conservative coupling) to $\pi/2$ (dissipative coupling) \cite{Stedman}.

\begin{table}[h!!!]
\noindent \begin{centering}
\begin{tabular}{|c|c|c|}
\hline 
Parameter & \multicolumn{2}{c|}{Value }\tabularnewline
\hline 
$\omega_s$ &  \multicolumn{2}{c|}{ $ 2 \pi \ 107.3 $ Hz $\simeq 674.5$ rad/s }\tabularnewline
\hline 
$\alpha$ & $4.3\times10^{-9}$ & $0.24$ rad/s \tabularnewline
\hline 
$\beta$ & $1.5\times 10^{-5}$ & $825$ rad/s \tabularnewline
\hline 
$r$ & $2\times10^{-7}$ & $11.1$ rad/s \tabularnewline
\hline 
\end{tabular}
\par\end{centering}

\caption{\label{RefPar} Typical values of Lamb parameters used in the simulations of G-Pisa dynamics with $G\sim3\cdot10^{-5}$. In the first column the units are adimensional Lamb units. To compare the order of magnitude of effects with $\omega_s,$ in the second column we report such parameters multiplied by the free spectral range $c/L$.}
\end{table}

\subsection{Steady state approximated solutions}

To provide suitable algorithms for parameter estimation, we study the steady state r\`egime of Eqs.(\ref{eq:GP}). By inspection of the right hand side of Eqs. (\ref{eq:GP}), one finds that the general steady state solutions are periodic. In particular, without backscattering ($r_{1,2}=0$), Eqs. (\ref{eq:GP}) exhibit steady state solutions of the type

\begin{equation}
\begin{cases}
I_{1}(t)\,&=\,\frac{\alpha_1}{\beta} \\
I_{2}(t)\,&=\,\frac{\alpha_2}{\beta} \label{eq:0ordsol} \\
\psi(t)\,&=\,\omega_s t \qquad,\\ 
\end{cases}
\end{equation} 
for $t \rightarrow \infty.$ In the presence of backscattering, the above solutions switch to periodic steady state solutions 
and exhibit oscillatory behaviors, and the backscattering can be treated as a perturbative 
sinusoidal forcing term. We can study the system oscillation around its unperturbed steady state by means of the time dependent
 perturbation theory \cite{goldstein}. To this aim we introduce the expansion 
parameter $\lambda,$ which is assumed to be of the same order of magnitude of $r_{1,2}$ and write:
\begin{equation}
\begin{cases}
\mathcal{I}_{1}(\lambda,t)\,&=\,\displaystyle{\sum_{k=0}^{\infty}\frac{\lambda^k }{k!}I_{1}^{(k)}(t)} \\
\mathcal{I}_{2}(\lambda,t)\,&=\,\displaystyle{\sum_{k=0}^{\infty}\frac{\lambda^k }{k!}I_{2}^{(k)}(t)} \\
\mathcal{\varPsi}(\lambda,t)\,&=\,\displaystyle{\sum_{k=0}^{\infty}\frac{\lambda^k}{k!}\psi^{(k)}(t)} \\ 
\end{cases}
\label{solutions}
\end{equation}

For $k = 0,$ substituting the latter into the Eqs.(\ref{eq:GP}), we recover the solution (\ref{eq:0ordsol}) with 
the positions $I_1^{(0)}(t) = \alpha_1 / \beta,$ $I_2^{(0)}(t) = \alpha_2 / \beta,$ {{$\psi^{(0)}(t) = \omega_s t$}}. The approximated solutions can be calculated iteratively from the series expansion in powers of $\lambda$ of (\ref{solutions}) into the dynamic of Eqs(\ref{eq:GP}). A second order approximations of the solutions reads:

\begin{equation}
\begin{cases}
\mathcal{I}_{1}(t) \! \! \! \! \! \! \!&\simeq \displaystyle{\frac{\alpha_1}{\beta }+2 r_2 \sqrt{\alpha_1 \alpha_2} \frac{ \alpha_1 \cos(\varepsilon + \omega_s t)+(\frac{\omega_s}{{c/L}}) \sin(\varepsilon +\omega_s t)}{\beta  \left(\alpha_1^2+{(\frac{\omega_s}{{c/L}})^2} \right) } -  2\frac{r_1 r_2({c/L})}{\beta  \omega_s} \sin(2 \varepsilon)} \\
\\
\mathcal{I}_{2}(t) \! \! \! \! \! \! \!&\simeq \displaystyle{\frac{\alpha_2}{\beta }+2 r_1 \sqrt{\alpha_1 \alpha_2} \frac{ \alpha_2 \cos(\varepsilon - \omega_s t)-(\frac{\omega_s}{{c/L}}) \sin(\varepsilon -\omega_s t)}{\beta  \left(\alpha_1^2+{(\frac{\omega_s}{{c/L}})^2} \right) } +2\frac{r_1 r_2({c/L})}{\beta  \omega_s} \sin(2 \varepsilon) }  \\
\\
\varPsi(t) \! \! \! \! \! \! \! &\simeq \displaystyle{(\omega_s  - \frac{2 r_1 r_2 ({c/L})^2 \cos(2 \varepsilon) } {\omega_s})t+ ({c/L})\frac{r_1 \sqrt{\frac{\alpha_1}{\alpha_2}} \cos( \varepsilon - \omega_s t) + r_2  \sqrt{\frac{\alpha_2}{\alpha_1}} \cos( \varepsilon + \omega_s t)  }{ \omega_s}}\ ,\\ 
\end{cases}
\label{eq:2ordsol}
\end{equation}
where we made the additional approximation of keeping the leading terms in $ \omega_s^n ,$ with {{$ n \leq 2$}}.  
Solutions (\ref{eq:2ordsol}) show a correction to the mean intensity level and pushing and pulling in the phase difference, 
as well as the presence of the first harmonic of $\omega_s$. 

\subsection{ \label{sect3}Parameter Estimation and Kalman filtering}

The Sagnac phase can be conveniently estimated by means of the Hilbert Transform (HT) of the interferogram $S(t)\simeq\sin\psi(t)$ which is routinely acquired during ring laser operation together with the monobeam intensities $I_{1}(t)$ and $I_{2}(t)$ \cite{Interferogram}. Since ring laser signals are sampled, we shift to the discrete time domain, and from here on we denote time dependent intensities and Sagnac phase as $\{I_{1}(n)\},$ $\{I_{2}(n)\}$ and $\{\psi(n)\}$  ($n\in\mathbb{N}$), respectively.

Perturbative solutions and the least squares methods provide the main tools for estimation procedure of the parameters $\alpha_{1,2},\ r_{1,2}$ and $\varepsilon.$ For sake of clarity, we re-parametrize the measured intensities and Sagnac phase in the form 
\begin{equation}
\begin{cases}
\mathcal{I}_{1}(t)\,=\, & I_{1}+i_{1} \sin\left(\omega t +\phi_{1}\right)\\
\mathcal{I}_{2}(t)\,=\, & I_{2}+i_{2} \sin\left(\omega t +\phi_{2}\right)\\ 
\mathcal{\varPsi}(t)\,=\,  &\omega t \\ 
\end{cases}
\label{EqId}
\end{equation}

as the quantities $I_{1,2}$, $i_{1,2}$ and $\phi_{1,2}$ can be readily estimated from  $\{I_{1,2}(n)\}$ and  $\{S(n)\}$. In fact, the mean intensities $I_{1,2}$ can be estimated by computing the sample average of $\{I_{1,2}(n)\}.$ Moreover,  the averaged modulation amplitudes $i_{1,2}$ and phase difference $\phi_{1}-\phi_{2}$ at the fundamental frequency can be estimated by means of a digital lock-in procedure, which calculates the ``in-phase'' and ``in-quadrature'' components of $\{I_{1,2}(n)\}.$ The reference complex signal for the digital lock-in is given by the HT of  $\{S(n)\}$. Averages are taken over a time interval where Lamb parameters remain fairly constant. The estimation procedure of Lamb parameters can be conveniently divided into two steps:

\begin{enumerate}
\item The first step is to estimate the phases of the intensities. From Eqs.(\ref{eq:2ordsol})  in the approximation { $\frac{\omega_s}{(c/L)}>>\alpha_{1,2}$} we have 

\[
\begin{cases}
\phi_1 = \varepsilon\\
\phi_2 = - \varepsilon \quad .\\ 
\end{cases}
\]

Thus we can immediately identify the backscattering angle from $\phi_{1,2}$ as

\begin{equation}
\widehat{\varepsilon}=\,\frac{\phi_{1}-\phi_{2}}{2}\quad ,
\end{equation}

where the hat symbol $\, \widehat{\,} \,$ denotes an identified parameter.

\item In the second step the remaining Lamb parameters are obtained by least squares methods. In fact, starting from the periodic steady state solutions (\ref{EqId}), we can form the squared residuals

\begin{align}
\Lambda & (\alpha_1,\alpha_2  ,r_1,r_2)  = \displaystyle{ \frac{2 \pi}{\omega} \times } \nonumber \\
& \times \displaystyle{\int^{2 \pi / \omega}_0 \Big \{ \dot{ \mathcal{I}_{1}}  - \frac{c}{L} \left[  \alpha_{1} \mathcal{I}_{1} - \beta \mathcal{I}_{1}^{2} + 2 r_{2} \sqrt{ \mathcal{I}_{1} \mathcal{I}_{2} } \cos(\omega t +\widehat{\varepsilon}) \right] \Big \}^2 + } \nonumber \\
& + \displaystyle{ \Big \{ \dot{ \mathcal{I}_{2}} - \frac{c}{L} \left[ \alpha_{2} \mathcal{I}_{2} - \beta \mathcal{I}_{2}^{2} + 2 r_{1} \sqrt{\mathcal{I}_{1} \mathcal{I}_{2}} \cos(\omega t -\widehat{\varepsilon})\right] \Big \}^2 \ dt }
\end{align}

averaged over a period $2\pi / \omega.$ Minimization of $\Lambda(\alpha_{1,2},r_{2,1})$  yelds the best linear estimate of $\alpha_{1,2}$ and $r_{1,2};$ from the conditions $\partial \Lambda / \partial \alpha_{1} = 0,$ $\partial \Lambda / \partial \alpha_{2} = 0$, 
$\partial \Lambda / \partial r_{2} = 0,$ and $\partial \Lambda / \partial r_{1} = 0,$  we get

\begin{eqnarray}
\widehat{\alpha_1}& = &\displaystyle{ \beta \left( I_1 + \frac{i_1^2}{4 I_1} \right) + \frac{i_1 i_2 \omega }{4 {{(c/L)}} I_2} \sin 2\widehat{\varepsilon}  } \\
\widehat{\alpha_2}&= &\displaystyle{ \beta \left( I_2 + \frac{i_2^2}{4 I_2}  \right) - \frac{i_1 i_2 \omega }{4 {(c/L)}  I_1} \sin 2\widehat{\varepsilon} } \\
\widehat{\, r_1 \,}& =  & \displaystyle{ \frac{i_2 \omega}{2 {(c/L)} \sqrt{I_1 I_2}} }\\
\widehat{\, r_2 \,} &=  & \displaystyle{ \frac{i_1 \omega}{2 {(c/L)} \sqrt{I_1 I_2}} } \quad,
\end{eqnarray}

\end{enumerate}

which fulfill the parameter estimation procedure via the second order approximation. It is worth noticing that one can increase the precision of the identified parameters by evaluating solutions of Eqs. (2) of higher order in $\lambda$ and increase their accuracy by increasing the averaging time span.

\section{\label{sect4} Simulation results for G-Pisa}

\begin{figure} 
\centering
\includegraphics[width=11cm]{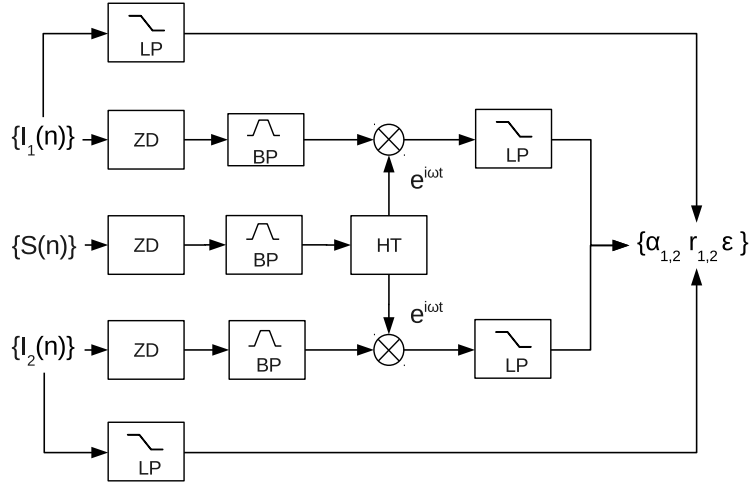}
\caption{Schematic of the parameter estimation procedure, where LP $\rightarrow$ lowpass Butterworth filter, BP $\rightarrow$ bandpass Butterworth filter, ZD $\rightarrow$ Zoom and Decimation routine, HT $\rightarrow$ Hilbert transform (see text).}
\label{fig:IdScheme}
\end{figure}

We briefly describe the specific implementation of the parameter estimation procedure for the G-Pisa ring laser, where the data are acquired at 
a sampling frequency of $5$ kHz ($T_s=200$ $\mu s$). To remove the oscillating component, intensity signals are low-pass filtered with a first 
order Butterworth filter with 1 Hz cutoff frequency.  The quantities $ I_{1,2} $ 
are estimated by averaging the decimated intensities over a time interval of  $10\ s$ (i.e. $5\times10^{4}$ samples). On the other side, to 
calculate the modulation $i_{1,2}$ and phases $\phi_{1,2}$, the intensities are first band-passed around the fundamental Sagnac band $[95\div125]$ Hz 
by means of a Butterworth filter, and decimated by a factor $2.$ The decimation procedure has been carried out by the tail recursive 
routine ``Zoom and Decimation of a factor $2^{n}$'' (ZD($n$)), where each iteration step is composed by a half band filter stage with 
discrete transfer function ${\displaystyle H(z)\,=\,\frac{z^{3}+2z^{2}+2z+2}{4z^{3}+2z}}$, followed by a downsampling by  $2$. 
The ZD($n$) procedure ensures a linear phase filter response at least for $n=3$ iterations, as no appreciable phase distortion 
was observed in simulated sinusoidal signals. The resulting data are then demodulated with a digital lock-in using as reference signal 
the discrete HT of the interferogram, and setting the integration time to 10 s. A schematic of the parameter estimation procedure is reported in Fig. \ref{fig:IdScheme}. In addition, the phase of the two monobeam oscillating components is determined by the discrete HT, and their difference is estimated by unwrapping the phase angle and taking its average over $10$ s. As a concluding remark on the parameter estimation procedure, we mention that the problem of filtering very long time series, has been solved by the ``overlap and save'' method \cite{mano}, which is an efficient algorithm for avoiding the boundary transients due to finite length of digital filters.

Reliability of the parameter estimation routine is tested by Monte Carlo simulation of the dynamics of Eqs.(\ref{eq:GP}) 
followed by the estimations of Lamb parameters from the simulated  time series 
of $I_{1,2}$ and $\psi$. We run $10^{4}$ simulations of the dynamics of Eqs.(\ref{eq:GP}) 
allowing $\alpha_{1,2}\ $ and $r_{1,2}$ to vary according to normal distributions with 
mean as in Tab. \ref{RefPar} and standard deviation equal to $10 \%$ of their means. 
In addition, $\beta$ is assumed constant, and $\varepsilon$ uniformly distributed 
in $[0, \pi/2)$. In each simulation, we have compared the numerical RK4 solution 
of Eqs. (\ref{eq:GP}) and approximated analytical solution (\ref{eq:2ordsol}) evaluated 
with the same Lamb parameters. We found that they are in a very good agreement, 
with means of the relative errors on $I_1(t),\ I_2(t) $ and $\psi(t)$ of $-6.4\times10^{-7}, \ -6.2\times10^{-7}$ and $-1.5\times10^{-5}$, 
and standard deviations of  $4.6\times10^{-6}, \ 4.6\times10^{-6}$ and $1.3\times10^{-6}$, respectively.

\begin{figure}
\centering
\includegraphics[width=14cm]{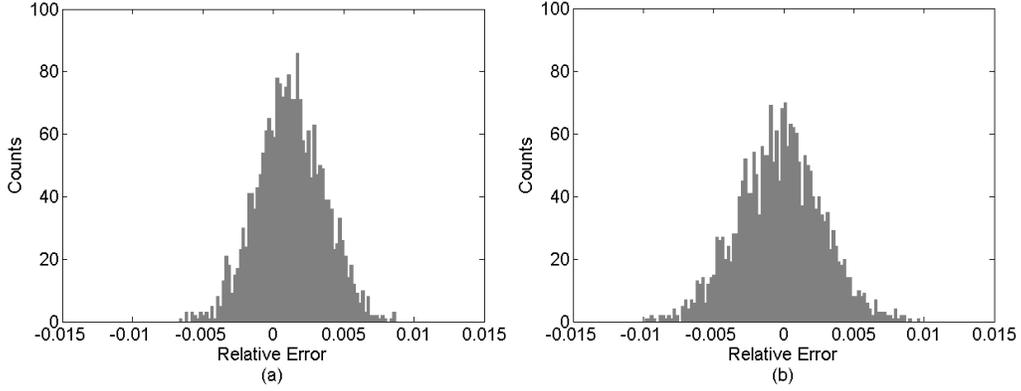}
\caption{\label{fig:IdResults1} Histograms of the relative errors $(\widehat{\alpha}_{1,2} - \alpha_{1,2})/ \alpha_{1,2}$ that affect the estimation of Gain minus losses parameters calculated with $2\times 10^4$ realizations of the ring laser dynamics. (a) Histogram relative to $\alpha_{1}:$ mean $1.4 \times 10^{-3}$ and  standard deviation $2.9 \times 10^{-3};$ (b) histogram relative to $\alpha_{2}:$ mean $-2.5 \times 10^{-4}$ and standard deviation $3.9 \times 10^{-3}$.}
\end{figure}

\begin{figure}
\centering
\includegraphics[width=14cm]{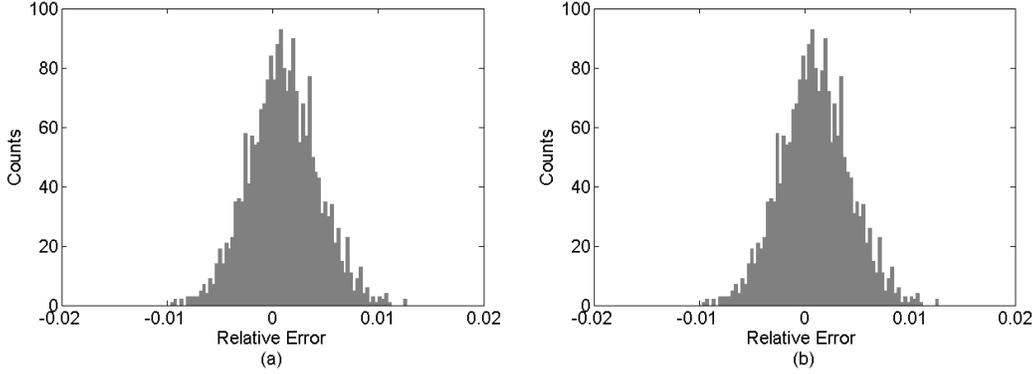}
\caption{ \label{fig:IdResults2} Histograms of the relative errors $(\widehat{r}_{1,2} - r_{1,2})/r_{1,2}$ that affect the estimation of backscattering coefficients calculated with $2\times 10^4$ realizations of the ring laser dynamics. (a) Histogram relative to $r_{1}:$ mean $1.1 \times 10^{-3}$ and  standard deviation $4.6 \times 10^{-3};$ (b)  histogram relative to $r_{2}:$ mean $1.3\times10^{-3}$ and standard deviation $3.2\times10^{-3}.$}
\end{figure}

\begin{figure}
\centering
\includegraphics[width=14cm]{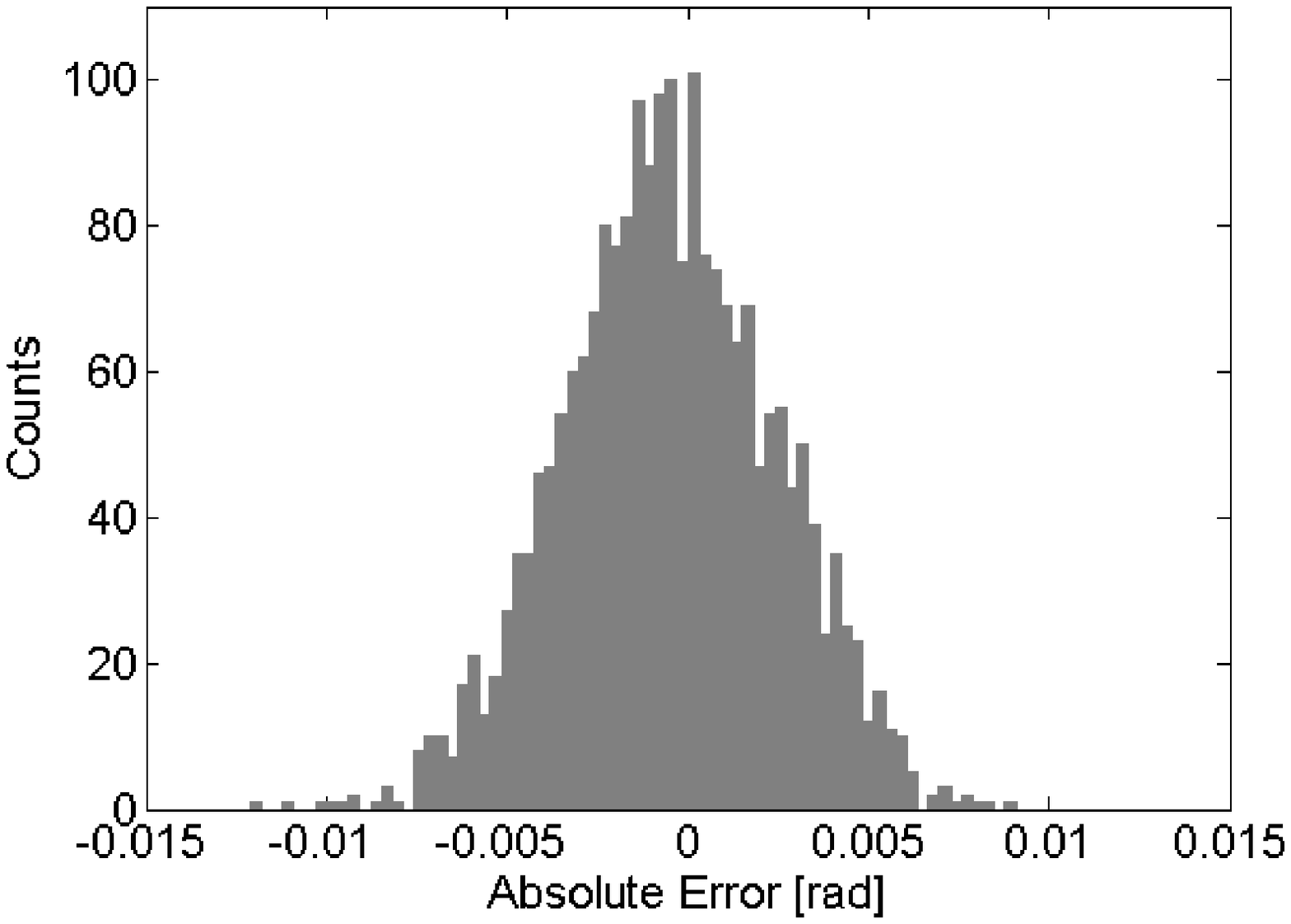}
\caption{ \label{fig:IdResults3} Histogram of the absolute errors $\widehat{\varepsilon} - \varepsilon$ that affect the estimation of backscattering phase calculated with $2 \times 10^4$ realizations of the ring laser dynamics; mean $-4.3\times 10^{-4}$ rad and standard deviation $2.8\times10^{-3}$ rad.}
\end{figure}

To numerically assess the performance of the parameter estimation procedure, we run a simulation 
of $6$ hours where $\alpha_{1,2},\ r_{1,2},\ \varepsilon$ and $\omega_{s}$ fluctuate following independent 
random walk processes with self-correlation time of $1$ hour. To reproduce the experimental behavior of a ring laser, 
the time drift of $\omega_{s},$ which mimics the effects of local tilts and rotations, is a factor of $5$ lower than the 
auto-correlation time of the other parameters. We superimposed to the simulated data an additive white noise, with SNR $= \ 10^2$ 
for the beam intensities and SNR $= \ 5 \times 10^3$ for the interferogram. 
Such order of magnitudes are routinely achieved in large ring laser \cite{wezzel} 
and in G-Pisa \cite{G-PISA}. The results we got are summarized in 
Fig. \ref{fig:IdResults1}, Fig. \ref{fig:IdResults2} and Fig. \ref{fig:IdResults3}.

The overall accuracy of the Lamb parameter estimation procedure is good, with a relative standard 
deviation of $3 \times 10^{-3}$ and $4 \times 10^{-3}$ in the estimation of $\alpha_{1,2},$ and $r_{1,2},$ respectively. 
The absolute error in the estimation of the backscattering phase is $3 \times 10^{-3}$ rad. The attained accuracy is not 
far from the lower bound $\sim 10^{-4}$ associated to the level 
of the observation noise of $\{I_{1}(n)\}$, $\{I_{2}(n)\}$ and $\{\psi(n)\}$.

\subsection{Estimation of $\omega_s$ by Kalman filtering}
Knowledge of the Lamb parameters $\widehat{\alpha}_{1,2}$ ,$\widehat{r}_{1,2}$ and $\widehat{\varepsilon}$, together with the $\beta$ parameter separately acquired, allow us to set up an Extended Kalman Filter \cite{EKF} for the estimation of the rotation rate $\widehat{\omega}_{s}$. 

The EKF state variables are the $\mathbb{R}^{3}$ vector $\mathbf{X}(t)\equiv\left[I_{1}(t),\ I_{2}(t),\ \psi(t)\right]^T$. The dynamics model is given by Eqs.(\ref{eq:GP}), with the addition of the model error as a zero mean, white, stochastic vector field $\mathbf{v}(t)$ with variance Var$[\mathbf{v}(t)]\equiv Q,$ where $Q$ is a $3\times3$ covariance matrix that accounts for the effects of unmodeled dynamics, for instance, identified parameter errors, calibration errors, and numerical integration inaccuracies. The EKF prediction step, which corresponds to the integration of Eqs.(\ref{eq:GP}) over the time interval $T_{s},$ is carried out using the RK4 Runge-Kutta routine.
 
In the discrete time domain, the model of the measurement process reads $\{ \mathbf{y}(n) \}\,=$ $\{\mathbf{X}(n)\} \ +\{ \mathbf{w}(n)\}$, where $\mathbf{w}(n)$ is zero mean, white, stochastic vector field (observation noise) with variance $Var[\mathbf{w}(n)]\equiv R,$ and $R$ is a $3\times3$ covariance matrix. In the standard experimental set up of ring lasers $\left[I_{1}(t),\ I_{2}(t),\ \psi(t)\right]$ are measured by independent sensors,  and so we can assume that $R$ is diagonal, with diagonal elements the observation noise variances $\sigma^2_{I_1},$ $\sigma^2_{I_2},$ $\sigma^2_{\psi}$ which can be conveniently calculated through the level of white noise in the power spectrum of $\{ \mathbf{y}(n) \}$.

The backscattering frequency is estimated from the filtered channels $\widehat{I}_{1,2}(n),\ $ $\widehat{\psi}(n)$, the identified parameters $\widehat{\alpha}_{1,2},\ \widehat{r}_{1,2}$, $\widehat{\varepsilon}$, and the exogenous parameter $\beta$ as 

\[
\widehat{\omega}_{BS}=\frac{c}{L}\left[\widehat{r}_{1}\sqrt{\frac{\widehat{I}_{1}}{\widehat{I}_{2}}}\sin(\widehat{\psi}-\widehat{\varepsilon})+\widehat{r}_{2}\sqrt{\frac{\widehat{I}_{2}}{\widehat{I}_{1}}}\sin(\widehat{\psi}+\widehat{\varepsilon})\right]\ 
\]

where, for simplicity, we have dropped the index $(n)$ from time series. The Sagnac frequency is then estimated from 
the difference $\widehat{\omega}_s = \dot{\widehat{\psi}}-\widehat{\omega}_{BS},$ where the numerical derivative of $\widehat{\psi}$ has been computed by the ``$5$ point method'' \cite{NumRec} designed to reject the derivative amplification of the noise.

\begin{figure}
\centering
\includegraphics[width=14cm]{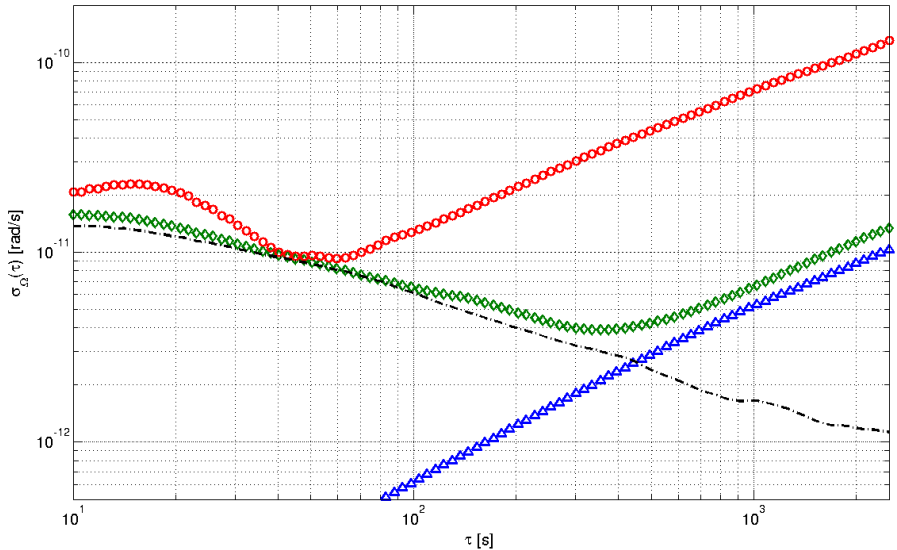}
\caption{\label{fig:EKFResultsSim} Allan standard deviation of the rotation rate estimated by AR(2) 
method (circles) and Extended Kalman Filter (open diamonds)  using  $2 \times 10^4$ seconds 
of the Monte Carlo simulation with random walk of Lamb Parameters. For comparison, we also plot 
the Allan standard deviation of the simulated rotational drift  (open triangles), and the Allan 
standard deviation of the EKF estimation after the subtraction of the rotational drift (dash dotted line). }
\end{figure}

The capability of the EKF in increasing the time stability 
and the resolution of the gyroscope has been tested with a $6$ hours simulation 
of the ring laser dynamics with parameter variations as in the parameter estimation tests. 
The results are summarized in Fig.\ref{fig:EKFResultsSim} where we compared the Allan variance 
of AR(2) and EKF frequency estimations. We conclude that, for this simulation with typical 
parameters of middle-size rings, the rotational resolution increases by a factor of $10$ while 
the minimum of the Allan standard deviation  shifts from $60$ s to $360$ s.

\section{\label{sect5} Application to real data}
To run the parameter estimation routine on the G-Pisa data, the photodetector signals $V_{1,2}(t)$ have been converted to dimensionless 
intensities $I_{1,2}(t)$ (Lamb units),
 using the following relation
\begin{equation}
I_{1,2}(t) = c_{Lamb} \frac{V_{1,2}(t) }{G_{ph} a_{eff} } \equiv c_{Lamb} P_{out\,1,2} , 
\end{equation}  
where: $G_{ph} = 10^9$ V$/$A is the photo-amplifiers gain, $a_{eff} = 0.4$A$/$W is the quantum efficiency of the 
photo-diodes, and $c_{Lamb} = 3.5 \times 10^6$ W$^{-1}$ is the calibration constant to 
Lamb units ( see Appendix \ref{sec:Lamb-Coefficient-Calculation}); $P_{out\,1,2}$ is the output power in Watt. 
The parameter estimation for G-Pisa  is completed by the acquisition and calibration of the $\beta$ parameter. 

\begin{figure}[h]
\centering
\includegraphics[width=14cm]{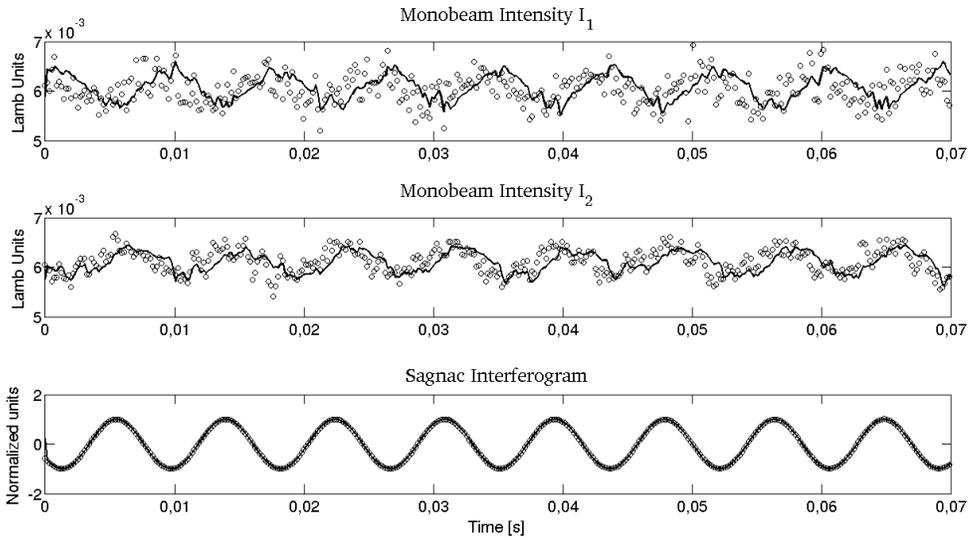}
\caption{Extended Kalman Filter estimation of $I_1(n),\ I_2(n) $ and $\sin{\psi(n)}$. Circles: G-Pisa raw data sampled at 5 kHz.
Continuous line: filter output.}
\label{timeseries}
\end{figure}

In Fig. \ref{timeseries} we show the comparison between the real time series measured on the G-Pisa ring laser and the 
calculated signal by the model after the parameter estimation according to the scheme in Fig. \ref{fig:IdScheme}.

\begin{figure}[h]
\centering
\includegraphics[width=14cm]{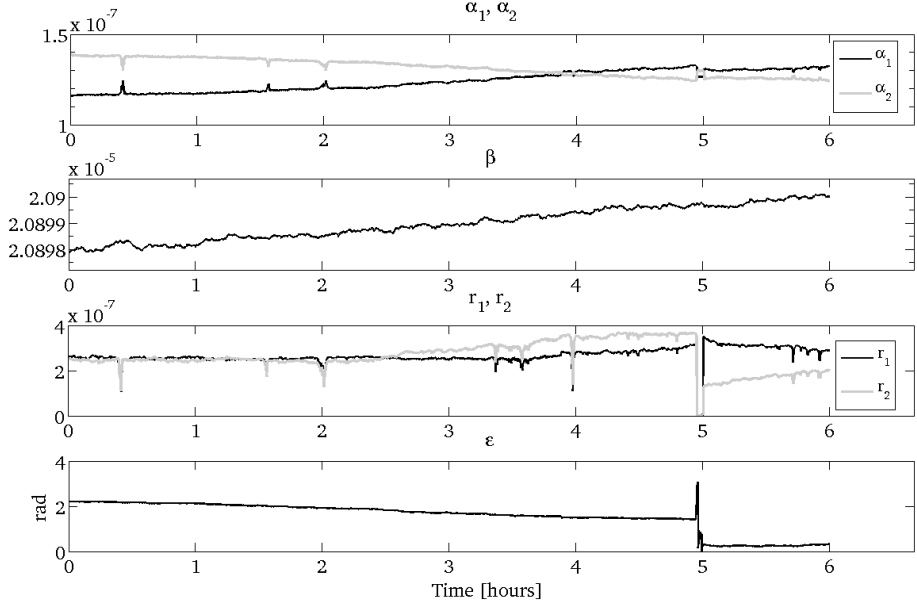}
\caption{Time series of  $\alpha_{1,2},$ $r_{1,2}$ and $\varepsilon$  and of $\beta$, respectively estimated and calibrated using 6 hours of experimental data of G-Pisa.}
\label{IDbadge}
\end{figure}

In Fig. \ref{IDbadge} we show the time series of the identified parameters for the G-Pisa ring laser and the calibrated 
parameter $\beta,$ using the routine described in Section \ref{sect3}.

\begin{figure}
\centering
\includegraphics[width=14cm]{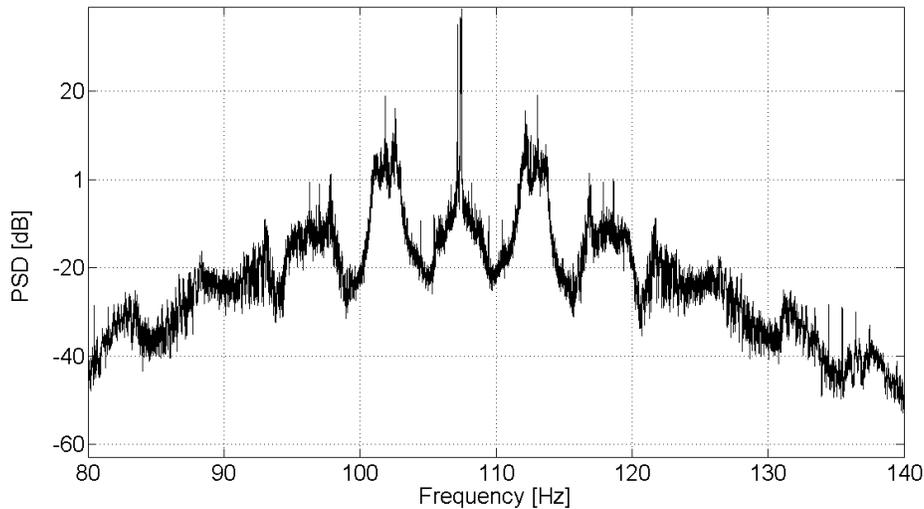}
\caption{\label{fig:psd} Power spectrum of the interferogram data around the Sagnac frequency $\sim 107.3$ Hz.}
\end{figure}

After the estimation of the G-Pisa parameters, we apply the EKF to the light intensities and to the interferogram (see \cite{Interferogram}). 
However, the implementation of the EKF requires an estimation of the covariance matrices $Q$ and $R$ of observation and model errors. 
Typically $Q$ and $R$ are considered as tuning parameters and set on the base of trial-and-error procedures. 
In fact, we started from an initial raw estimation for the diagonal elements of $Q$ and $R$ using simulations and power spectra of 
$I_{1,2}$ and $\psi,$ respectively. Then we tuned these values searching for the minimum of the Allan variance of 
$\omega_s$ and came to $Q = \rm{diag}(10^{-8},10^{-8},10^{-10})$ and $R = \rm{diag}(10^{-8},10^{-8},10^{-8})$. 
The performance of parameter estimation and EKF were limited by the environmental conditions of G-Pisa, e.g. local tilts and spurious 
rotations induced by the granite slab that support the instrument and some electronic disturbances, as it can be seen in Fig. \ref{fig:psd}, 
where we report the power spectrum of  $\{S(n)\}$.  

\begin{figure}[h]
\centering
\includegraphics[width=14cm]{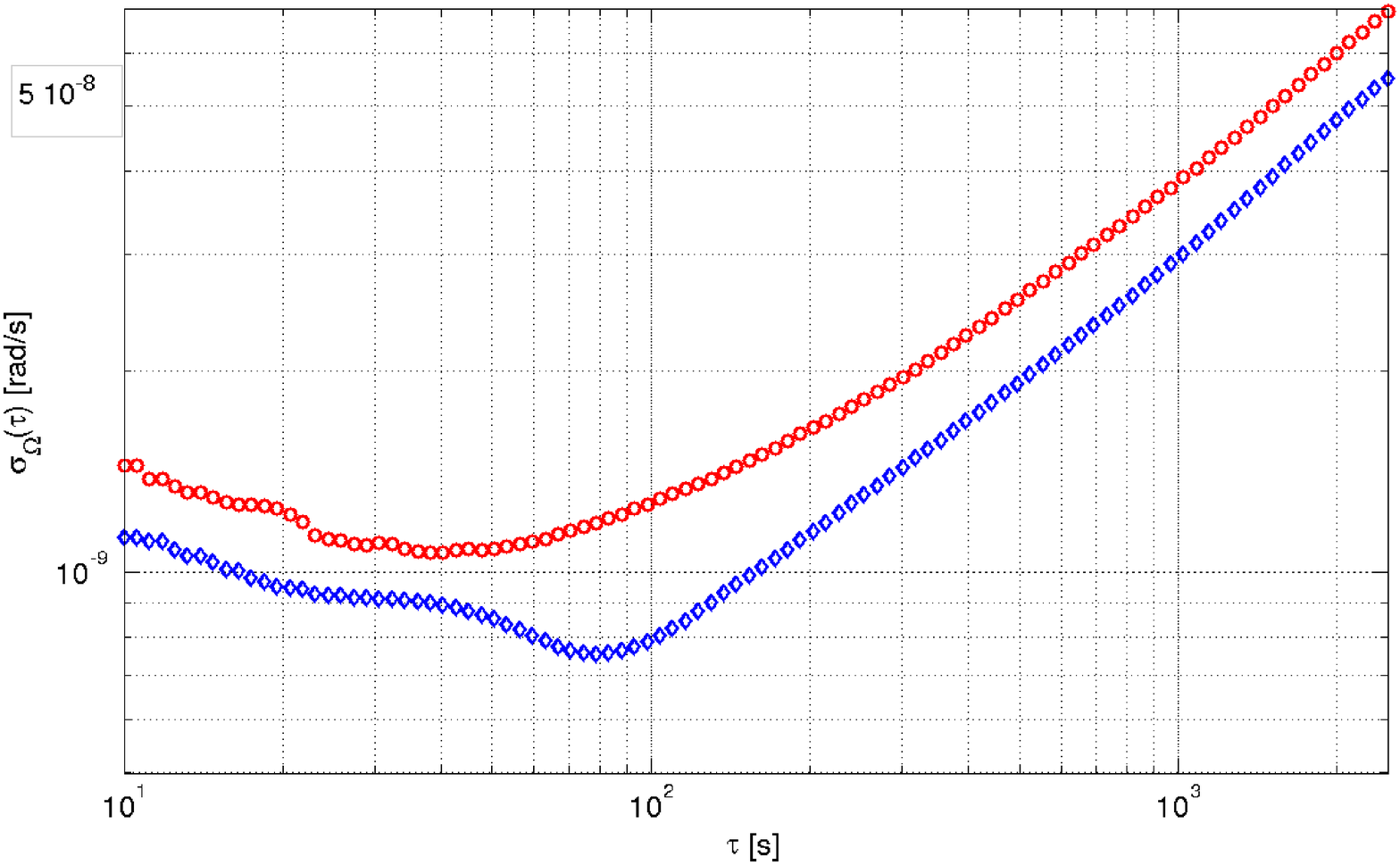}
\caption{\label{fig:EKFResultsXP} Allan standard deviation of the rotation rate estimated by AR(2) method (circles) and Extended Kalman Filter (open diamonds) using  $2 \times 10^4$ seconds of experimental data of G-Pisa.}
\end{figure}

In figure \ref{fig:EKFResultsXP} 
we report the Allan standard deviation of the Sagnac frequency estimated with AR(2) and EKF. 
An increase of a factor of $1.5$ in rotation--rate resolution and of a factor of $2$ in the time stability is observed. 

\section{\label{sect6} Conclusions}
A full model of the ring laser dynamics has been studied and applied to the estimation and removal 
of the long term drift in the laser parameters. The proposed data processing technique is based on the estimation of the parameters appearing in the Lamb equations for a $\rm{He}$-$\rm{(^{20}Ne + ^{22}Ne)}$ ring laser. 
Results of Monte Carlo simulations supported the viability of the parameter estimation, and yielded a 
relative estimation error of the order of $3 \times10^{-3}$ for $\alpha_{1,2}$ and $r_{1,2}$ and an estimation error $\sim 10^{-3}$~rad for
 $\varepsilon$. The accurate estimation of the Lamb parameters allows for the application of the Kalman Filter for the estimation of the rotation rate from the Sagnac frequency. Simulations showed a significant improvement in the frequency estimation by EKF compared to AR(2) method. Preliminary results on data from the ring laser prototype G-Pisa, presently strongly affected by local rotational noise, make us confident about the reliability of our approach in the presence of unmodeled experimental noise and calibration errors. Our approach 
can be further improved, and different mathematical tools can be used. For instance, the numerical integration method RK-4 can be substituted by geometrical 
integrators, or modified in conservative routines \cite{GeoMeth}. EKF can be also improved by increasing the state dimension, 
modifying the observation model and the estimation of $\omega_s.$  
As a final remark, we note that ring lasers achieved world record in rotation sensitivity, accuracy and time stability with sophisticated hardware, 
accurate selection of the working point of the He-Ne laser, despite a very basic off-line analysis. 
In this paper, we have shown that the parameters of the ring laser dynamics can be identified and their 
effects on resolution and time stability removed notwithstanding the system non linearities. 
We think that data analysis will cooperate more and more with ring laser hardware in pushing the 
resolution and the time stability of ring lasers beyond the current limitations.

\appendix

\section{Lamb Coefficients calculation\label{sec:Lamb-Coefficient-Calculation}}

The two intensities $I_{1}$ and $I_{2}$ in Eqs.(\ref{eq:fond1}) are expressed in Lamb
Units 
\[
I_{1,2}\,=\,\frac{|\mu_{ab}|^{2}(\gamma_a + \gamma_b)}{4\hbar^{2}\gamma_{a}\gamma_{b}\gamma_{ab}}E_{1,2}^{2}\,=\,\frac{|\mu_{ab}|^{2}(\gamma_a + \gamma_b)}{4\hbar^{2}\gamma_{a}\gamma_{b}\gamma_{ab}}\cdot\frac{P_{out\,1,2}}{2c\epsilon_{0}s_{b}T }\,\equiv \, c_{Lamb} P_{out\,1,2} 
\]

where 

\[
\mu_{ab}=\sqrt{\pi\epsilon_{0}\frac{\lambda^{3}}{(2\pi)^{3}}\hbar A_{ik}}\ ,
\]
is the electric-dipole matrix element between the laser states $a=3s^2$ and $b=2p^4$ (i.e. the upper and the lower of the laser energy levels), 
$\gamma_{a}$ and $\gamma_{b} $ are the decay rates in Paschen notation, $\eta=\gamma_{ab}/ \Gamma_{D}$ is the ratio between the homogeneous and 
the Doppler broadening of the laser transition. Here $\Gamma_{D} = \sqrt{2 k_b T_p / m_{ne}}/ \lambda,$ where $k_b$ is the Boltzmann 
constant, $m_{ne}$ is the atomic mass of Neon and $T_p$ is the plasma temperature. $A_{ik}$ is the radiative decay rate between the laser 
levels, $E_{1,2}$ are the electric fields amplitudes, $P_{out\,1,2}$ are the output powers, $\epsilon_{0}$ is the dielectric constant of 
vacuum, $s_b$ is the area of the transverse section of the beam, $T$ is the transmission coefficient of the mirror, and $\hbar$ is the 
reduced Plank constant. The table \ref{laserpar}
contains the reference values of the above quantities for a Doppler broadened active medium 
in presence of collision, according to refs. \cite{Lamb} and \cite{Smith}. 

\begin{table}[h!!!]
\noindent \begin{centering}
\begin{tabular}{|c|c|}
\hline 
$s_b$ & $18 \cdot 10^{-6}\ \rm{m^{2}}$\tabularnewline
\hline 
$P_{out}$ & $1-10\ \rm{nW}$\tabularnewline
\hline 
$p$ & $5.25\ \rm{mbar}$ \tabularnewline
\hline 
$\gamma_{a}$ & $12\ \rm{MHz}$\tabularnewline
\hline 
$\gamma_{b}$ & $127\ \rm{MHz}$\tabularnewline
\hline 
$\gamma_{ab}$ & $234\ \rm{MHz}$\tabularnewline
\hline 
$T_{p}$ & $450\ \rm{K}$\tabularnewline
\hline 
$\mu_{ab}$ & $3.2\cdot10^{-30}\rm{C\, m}$\tabularnewline
\hline 
\end{tabular}
\par\end{centering}
\caption{ \label{laserpar} The G-Pisa laser parameters.}
\end{table}

The coefficients in Eq.(\ref{eq:fond1}) can be calculated by means of the plasma dispersion function, which is the function associated to the broadening profile of the laser transition

\begin{equation}
 {\displaystyle Z(\xi_{1,2})\,=\,2i\int_{0}^{\infty}e^{-x^{2}-2\eta x-2i\xi_{1,2}x}dx \quad ,}
\label{eq:DispFun}
\end{equation}

where $\xi_{1,2}=(\omega_{1,2}-\omega_{0})/\Gamma_{D}$ is the detuning, normalized to the Doppler width, from the transition center for the beams $1$ and $2.$  The independent variables $\xi_{1,2}$ are in correlation with temperature and pressure inside the cavity. For an active medium composed by a gas mixture of two isotopes, one has to account for $\xi$ and $\xi'$, $\eta$ and $\eta'$, $\Gamma_D$ and $\Gamma_D'$. Here the unprimed and primed symbols refer to the $Ne$ isotopes $20$ and $22,$ respectively. In the Doppler limit $\eta \ll 1$ and $\eta' \ll 1,$ which is common for middle or large size He-Ne rings, $Z(\xi)$ is usually approximated as

\begin{eqnarray*}
Z_{I}(\xi) & \simeq & \sqrt{\pi}e^{-\xi^{2}}-2\eta\\
Z_{R}(\xi) & \simeq & -2\xi e^{-\xi^{2}} \quad ,
\end{eqnarray*}

where the pedices $I$ and $R$ stands for imaginary and real part, respectively. Within the above approximations the Lamb coefficients have the following expressions

\begin{align}
\alpha_{1,2}\,= & \frac{G}{Z_{I}(0)} \left[ kZ_{I}(\xi_{1,2})+k'Z_{I}(\xi'_{1,2}) \right]\,-\,\mu_{1,2}\nonumber \\
\beta_{1,2}\,= & \alpha_{1,2}\,+\,\mu_{1,2}\nonumber \\
\sigma_{1,2}\,= & \frac{f_{0}}{2} \frac{G}{Z_{I}(0)} \left[ kZ_{R}(\xi_{1,2})+k'Z_{R}(\xi'_{1,2})\right] \label{eq:Lamb}\\
\theta_{1 2}\,= & \frac{\Gamma G}{Z_{I}(0)} \left[ k\frac{Z_{I}(\xi_{1,2})}{1+(\xi_{m}/\eta)^{2}}+k'\frac{Z_{I}(\xi'_{1,2})}{1+(\xi'_{m}/\eta)^{2}}\right] \nonumber \\
\tau_{1 2}\,= & \frac{\Gamma f_{0}}{2} \frac{G}{Z_{I}(0)} \left[ k\frac{Z_{I}(\xi_{1,2})\xi_{m} /  \eta }{1+(\xi_{m}/\eta)^{2}}+k'\frac{Z_{I}(\xi'_{1,2})\xi'_{m} /  \eta}{1+(\xi'_{m}/\eta)^{2}}\right] \ , \nonumber 
\end{align}

where $G$ is the laser gain, $\Gamma=(\gamma_a+\gamma_b)/(2 \gamma_{ab}),$ $\xi_m = (\xi_1 +\xi_2)/2,$ $\xi'_m = (\xi'_1 +\xi'_2)/2,$ $\mu_{1,2}$ are the mirror losses experienced by each beam, $k$ and $k'$ are the fractional amount of isotopes in the gas mixture. The equation for $\theta_{2 1}$ and $\tau_{2 1}$ are obtained from the expression of  $\theta_{1 2}$ and $\tau_{1 2}$ by permuting the subscripts 1 and 2. Substituting $\xi_{1,2} \simeq 0.47,$ $\xi_{1,2}' \simeq -0.49,$ $k=k'=0.5,$ $G \simeq 3 \times  10^{-5},$ $\eta \simeq 0.25,$ and $\mu_{1,2} \simeq 1,48 \times 10^{-5}$ in  Eqs. (\ref{eq:Lamb})  we get for the closed loop operation of G-Pisa $\beta_1 - \beta_2 \simeq 10^{-14},$ $\theta_{12,21} \simeq 8 \times 10^{-8},$ $\tau_{12,21} \simeq 10^{-1},$ and the values in Table \ref{RefPar}.


\begin{thebibliography}{References}
\bibitem{Interferogram} By definition, the interferogram of the two counter-propagating 
beams is given by $S(t)=I_{1}(t)+I_{2}(t)-2\sqrt{I_{1}(t)I_{2}(t)}\sin(\psi(t)).$ However, to estimate $\sin(\psi(t))$ directly from $S(t)$, the linear trend  $I_{1}(t)+I_{2}(t)$ is removed, and the energy 
$I_{1}(t)I_{2}(t) $ is normalized to 1 over time intervals which usually correspond to thousands of cycles.
\end{thebibliography}
\end{document}